\documentclass[12pt,preprint]{aastex}

\usepackage{graphicx}
\usepackage{mathrsfs}
\usepackage{amsmath}
\usepackage{epsfig}

\def\iso#1#2{\mbox{${}^{#2}{\rm #1}$}}

\def\c1#1{\iso{C}{1#1}}
\def\n1#1{\iso{N}{1#1}}
\def\o1#1{\iso{O}{1#1}}

\newcommand{\be}{\begin{equation}}
\newcommand{\ee}{\end{equation}}

\def\beq#1\eeq{\begin{equation}#1\end{equation}}
\def\beqar#1\eeqar{\begin{eqnarray}#1\end{eqnarray}}

\def\la{\mathrel{\mathpalette\fun <}}
\def\ga{\mathrel{\mathpalette\fun >}}
\def\fun#1#2{\lower3.6pt\vbox{\baselineskip0pt\lineskip.9pt
  \ialign{$\mathsurround=0pt#1\hfil##\hfil$\crcr#2\crcr\sim\crcr}}}
\def\ie{{\it i.e.},~}
\def\eg{{\it e.g.},~}

\def\yd{$y_{\rm D}$}
\def\ydp{$y_{\rm DP}$}

\def\ydism{$y_{\rm D,ISM}$}

\def\ydlb{$y_{\rm D,LB}$}

\def\ydmax{$y_{\rm D,max}$}
\def\ydmin{$y_{\rm D,min}$}

\def\hi{H\thinspace{$\scriptstyle{\rm I}$}}
\def\hii{H\thinspace{$\scriptstyle{\rm II}$}}
\def\di{D\thinspace{$\scriptstyle{\rm I}$}}

\def\4he{$^4$He}

\makeatletter

\newcommand{\Rmnum}[1]{\expandafter\@slowromancap\romannumeral #1@}
\makeatother

\def\apj{ApJ,~}
\def\apjl{ApJL,~}
\def\apjs{ApJS,~}
\def\aap{A\&A,~}

\begin{document}

\title {The Deuterium Abundance in the Local Interstellar Medium}

\author{Tijana Prodanovi\'{c}\altaffilmark{1}
\altaffiltext{1}{E-mail: prodanvc@df.uns.ac.rs}
\affil{Department of Physics, University of Novi Sad \\
Trg Dositeja Obradovi\'{c}a 4, 21000 Novi Sad, Serbia}}

\author{Gary Steigman\altaffilmark{2}
\altaffiltext{2}{E-mail: steigman@mps.ohio-state.edu}
\affil{Department of Physics and Astronomy, Ohio State University \\
191 W. Woodruff Ave., Columbus OH 43210-1117, USA}}

\author{Brian D. Fields\altaffilmark{3}
\altaffiltext{3}{E-mail: bdfields@illinois.edu}
\affil{Department of Astronomy, University of Illinois \\
1002 W. Green St., Urbana IL 61801, USA}}

\begin{abstract}

As the Galaxy evolves, the abundance of deuterium in the 
interstellar medium (ISM) decreases from its primordial value: 
deuterium is ``astrated".  The deuterium astration factor, $f_{\rm 
D}$, the ratio of the primordial D abundance (the D to H ratio by 
number) to the ISM D abundance, is determined by the competition 
between stellar destruction and infall, providing a constraint on 
models of the chemical evolution of the Galaxy.  Although conventional 
wisdom suggests that the local ISM (i.e., within $\sim 1-2$~kpc of 
the Sun) should be well mixed and homogenized on timescales short 
compared to the chemical evolution timescale, the data reveal gas 
phase variations in the deuterium, iron, and other metal abundances 
as large as factors of $\sim 4-5$ or more, complicating the estimate 
of the ``true" ISM D abundance and of the deuterium astration factor.  
Here, assuming that the variations in the observationally inferred 
ISM D abundances result entirely from the depletion of D onto dust, 
rather than from unmixed accretion of nearly primordial material,
a model-independent, Bayesian approach is used to determine the 
undepleted abundance of deuterium in the ISM (or, a lower limit 
to it).  We find the best estimate for the undepleted, ISM 
deuterium abundance to be (D/H)$_{\rm ISM}\geq (2.0 \pm 0.1) 
\times 10^{-5}$.  This result is used to provide an estimate of 
(or, an upper bound to) the deuterium astration factor, $f_{\rm D} 
\equiv$~(D/H)$_{\rm P}/$(D/H)$_{\rm ISM} \leq 1.4 \pm 0.1$.  

\end{abstract}

\maketitle


\section{Introduction}

Deuterium is created in an astrophysically interesting abundance 
only during big bang nucleosynthesis (BBN) \citep{boes,steigman07}, 
after which, in the post-BBN Universe, its abundance ($y_{\rm D} \equiv 
10^5 (\rm D/H)$) decreases monotonically due to the processing of gas 
through succeeding generations of stars where deuterium is completely 
destroyed \citep{els,pf03}. Consequently, deuterium plays a special role 
in cosmology, nuclear astrophysics, and in Galactic chemical evolution
\citep{ytsso,boes,st92,st95,vangioni,tosi96,schramm,romano06,srt,steigman07,pf08}.  
Its relatively simple Galactic evolution permits us to use deuterium to 
determine the fraction of interstellar gas that has been processed through 
stars \citep{st92,st95}.  By comparing the primordial and Galactic deuterium 
abundances one can learn about and discriminate among different Galactic 
chemical evolution models \citep{vangioni,tosi96,romano06,srt,pf08}.  Together 
with non-BBN constraints on the baryon (nucleon) density from observations of 
the cosmic microwave background \citep{spergel,Dunkley09,komatsu09,komatsu10}, 
the primordial deuterium abundance, \ydp, is predicted by BBN 
\citep{cyburt03,coc,steigman07,vs,cyburt08}.  The predicted abundance is 
in excellent agreement with observations of deuterium in high-redshift, 
low-metallicity QSO absorption line systems (QSOALS) \citep{omeara,pettini}.  

\begin{figure}[t]
\epsscale{0.4}
\plotone{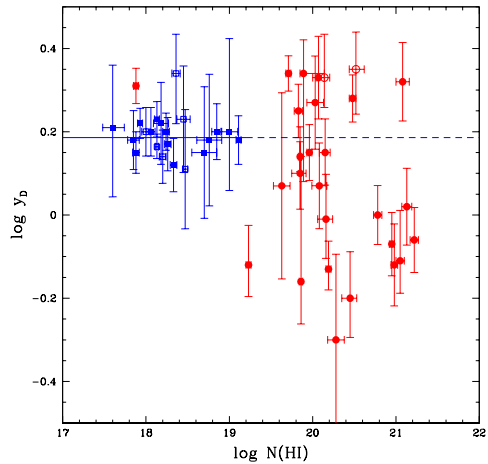}
\caption{The logs of the deuterium abundances versus the 
logs of the \hi~column densities [cm$^{-2}$] for the 49 FUSE 
LOS (see the text).  The filled symbols are for the 41 LOS 
which have iron abundance data, while the open symbols are 
for the 8 LOS which lack iron abundances.  The squares (blue) 
are for the LOS within the Local Bubble (LB) and the circles 
(red) are for the non-LB (nLB) LOS.  The solid line is at 
the mean D abundance for the LB LOS (log $y_{\rm D,LB} = 
0.19$); the dashed line is its extension to the nLB LOS.}
\label{fig:lgyvslgh}
\end{figure}

Since deuterium is destroyed in the Galaxy as gas is cycled through 
stars, (D/H)$_{\rm ISM} \leq$ (D/H)$_{\rm P}$.  However, all successful 
chemical evolution models require infall to the disk of the Galaxy of 
unprocessed (or, nearly unprocessed) gas \citep{tosi96,pf08}, and such 
deuterium-rich (and metal-poor) gas would raise the ISM D/H ratio 
closer to the primordial value.  As a result, comparison of the 
primordial and ISM deuterium abundances provides a constraint on 
infall and, therefore, on chemical evolution models.  Observations 
over the past decade and more of the deuterium abundance in the 
relatively local interstellar medium (ISM) reveal an unexpectedly 
large scatter in D/H \citep{jenkins,sonneborn,hebrard,hoopes}, 
challenging the conventional wisdom of a well mixed ISM.  For example, 
as shown in Figure~\ref{fig:lgyvslgh}, the absorption line measurements 
from the {\it Far Ultraviolet Spectroscopic Explorer} (FUSE) reveal 
variations of a factor of $\sim 4$ ($0.5 \la y_{\rm D,ISM} \equiv 
10^{5}$(D/H)$_{\rm ISM} \la 2.2$) in the {\em gas-phase} D/H ratios 
over lines of sight (LOS) to background stars within $\sim 1-2$ kpc 
of the Sun.  Moreover, the variations in the observed, gas phase 
D/H abundances are found to correlate {\em positively} with the 
abundances of refractory elements such as Ti \citep{prochaska,lallement08}, 
and a similar correlation is also found between D and other metals 
such as Fe and O \citep{linsky,srt}.  However, this positive correlation 
between the abundances of D and the metals is opposite to the trend 
expected from stellar nucleosynthesis (D decreasing as the metals 
increase).  Motivated by the observed correlations and by the very 
large spread in the D/H ratios inferred from the FUSE and other data 
sets, it has been proposed that the large variations in {\em gas-phase} 
D/H may be due to the depletion of gas phase deuterium onto dust 
grains \citep{jura,draine04,draine06}.  If this is the case, then the 
FUSE (and other) gas-phase absorption-line measurements reveal 
that depletion has not been well mixed (homogenized) in the ISM 
and, the data may only provide a {\em lower} limit to the {\em true} 
(undepleted) ISM deuterium abundance and, therefore, only an 
{\em upper} limit to the deuterium astration factor, $f_{\rm D} 
\equiv y_{\rm DP}/y_{\rm D,ISM}$.  

In Figure~\ref{fig:lgyvslgh} the logs of the deuterium abundances 
(log~$y_{\rm D} \equiv 5~+~$logN(\di) $-$ logN(\hi)) are shown as 
a function of the logs of the \hi~column densities for the 49 LOS 
with \di~column densities from Table 2 of \citet{linsky}, supplemented
by data from \cite{oliveira2006} and \cite{dupuis2009}.  Of these 49 
LOS, 41 have iron abundance measurements (filled symbols); 21 of 
the 49 LOS are within the Local Bubble (LB; see \citet{linsky} for 
a discussion of the LB); the remaining 28 non-Local Bubble (nLB) 
LOS are toward stars beyond the LB (see \S2.1).  

The unexpectedly large spread among the observationally inferred 
ISM D abundances complicates any estimate of \ydism.  Recognizing 
this point, \citet{linsky} chose for their estimate of a {\em lower 
bound} to the true (undepleted) ISM deuterium abundance the mean 
of the five {\em highest} D/H ratios finding \ydism~$\geq 2.17 \pm 
0.17$ or, when including corrections (see the discussion in \S2.1) 
for N(\hi) and N(\di) for those lines of sight outside of the Local 
Bubble, \ydism~$\geq 2.31 \pm 0.24$.  These estimates are quite close 
to the {\it lower bound} to the primordial abundance estimated from 
the QSOALS, $y_{\rm DP} = 2.82^{+0.20}_{-0.19}$ \citep{pettini}\footnote{
Since this estimate of \ydp~relies on only 7 high redshift, low 
metallicity LOS and, since the dispersion in deuterium abundances 
among them is unexpectedly large, some prefer to adopt a so-called 
``WMAP D abundance".  WMAP does not observe deuterium.  Rather, the 
WMAP-determined baryon density parameter may be used in a BBN code 
to {\it predict} the relic D abundance.  If the \cite{komatsu10} 
estimate of the baryon density is adopted, the BBN predicted 
primordial D abundance is \ydp~$= 2.5 \pm 0.1$.  However, the 
WMAP collaboration also provides an estimate of the effective 
number of neutrinos, \citep{komatsu10} which, when used along 
with their baryon density estimate in a BBN code, leads to a 
{\it different} predicted primordial D abundance \ydp~$= 3.0 
\pm 0.4$.  The difference between these two predictions reflects 
the difference between the standard model effective number of 
neutrinos expected and the WMAP observed value, which is within 
$\sim 1.5\sigma$ of expectations.  So, which, if either, ``WMAP 
D abundance" is preferred?  Here, we compare observations to 
observations, not to model-dependent predictions and, we adopt 
the \cite{pettini} value for \ydp.}, suggesting a small upper 
bound to the D astration factor, $f_{\rm D} \leq 1.30^{+0.14}_
{-0.13}$ or, an even smaller value, $f_{\rm D} \leq 1.22 \pm 
0.15$ for the LB-corrected, nLB deuterium abundances.  To account 
for such a high ISM deuterium abundance and such a small D 
astration factor, a very high infall rate of pristine material 
would be needed, challenging many Galactic chemical evolution 
models \citep{romano06,pf08}.  By limiting themselves to the five 
highest D abundances, \citet{linsky} ignore the lower deuterium 
abundances along many more LOS which are consistent with them 
within the errors, potentially biasing their estimates of 
\ydism~and of $f_{\rm D}$.  

In an attempt to address this issue, \citet{srt} (SRT), used the 18 highest 
D/H ratios from the FUSE data (see Table 3 of \citet{linsky}), finding an 
ISM D abundance of $y_{\rm D,ISM} = 1.88 \pm 0.11$, corresponding to a 
D-astration factor $f_{\rm D} \leq 1.50^{+0.14}_{-0.13}$, consistent 
with at least some of the otherwise successful chemical evolution 
models discussed in SRT.  In fact, the data in Table 2 of \citet{linsky} 
reveal that there are 19 LOS with central values of log \yd~$\geq 0.20$.  
The weighted mean (along with the error in the mean) for these 19 D 
abundances is log $y_{\rm D,19} = 0.26 \pm 0.01$, corresponding to 
$y_{\rm D,19} = 1.8 \pm 0.1$.  For these 19 LOS the reduced $\chi^{2} 
= 0.85$, confirms that the weighted mean provides a good description 
of their D abundances.  As more LOS with lower D abundances are added, 
the weighted mean decreases, but the reduced $\chi^{2}$ increases, so 
that \ydism~$\ga 1.8 \pm 0.1$ is likely a robust {\it lower} bound to the 
ISM D abundance and, $f_{\rm D} \la 1.5 \pm 0.1$ (log $f_{\rm D} \la 
0.19 \pm 0.03$) is a robust {\it upper} bound to the deuterium astration 
factor.

Surely, there must be a better way to find a reliable estimate of 
the maximum value of the deuterium abundance in the local ISM while 
accounting for the observational errors in the individual D abundance 
determinations.  In \S 2 we describe a Bayesian analysis designed to 
find the best estimate of the maximum, gas phase (undepleted), ISM 
deuterium abundance, \ydmax, from data with non-negligible errors and 
we apply it to the FUSE data set.  On the assumption that the spread 
in the observed D abundances is the result of incompletely homogenized 
dust depletion, \ydism~$\geq$ \ydmax~and $f_{\rm D} \equiv y_{\rm 
DP}/y_{\rm D,ISM} \leq y_{\rm DP}/y_{\rm D,max}$.  Our results are 
summarized and our conclusions presented in \S 3.  As already noted, 
the dust depletion hypothesis suggests that there should be a correlation 
between the D and Fe abundances. Although this trend appears to be 
present at some level \citep{linsky}, a deeper analysis suggests that 
the simplest interpretation of the depletion hypothesis may need to 
be modified~\citep{srt}. For example, since D is likely attached to the
mantles of dust grains \citep{tielens}, while Fe is mostly contained in 
the cores of the grains, D will be removed from grains more easily than 
Fe. Thus, although strong shocks may destroy grains completely, weak 
shocks may only remove D from grains while Fe might stay locked within 
their cores \citep{draine79}.  This effect then may explain the scatter 
observed in the relation between the D and Fe depletions.  Thus, in 
a follow-up paper in preparation, the consistency of the depletion 
hypothesis with the FUSE data for D and Fe (and O) is investigated, 
and some possible modifications to its simplest version are explored 
\citep{sp}.

\section{A Bayesian Analysis Of The ISM D Abundances}

To avoid imposing any prior prejudice on which LOS should be 
included and which excluded in our analysis or, on how the 
observed D abundances may, or may not correlate with iron 
(or other metals), we adopt a statistical, model-independent 
method for determining the undepleted Galactic deuterium 
abundance.  Our approach follows closely the model-independent 
Bayesian analysis developed by \citet{hogan} to determine the 
primordial helium abundance.  It is useful to recall the problem 
\citet{hogan} addressed and how they solved it.  Helium-4 is 
produced abundantly during BBN and in the post-BBN Universe 
its abundance is enhanced by stellar produced \4he.  As a result, 
the \4he mass fraction, Y$_{\rm P}$, is expected to increase from 
its primordial value, Y$_{\rm P}$, along with the metallicity, Z.  
Extrapolation of the observed Y versus Z relation to zero metallicity 
results in an estimate of Y$_{\rm min} =$ Y$_{\rm P}$.  The problem 
for \4he is that the form of the Y$_{\rm P}$ versus Z relation is 
a priori unknown and, may even differ from object to object (low 
metallicity, extragalactic \hii~regions).  This, in combination 
with the errors in the observed values of Y$_{\rm P}$ and Z, 
complicates the derivation of Y$_{\rm P}$ from the data.  The 
challenge confronting \citet{hogan} was to identify Y$_{\rm min}$.  
The Bayesian approach they developed \citep{hogan} is described 
below.  If, indeed, the observed spread in D abundances results 
from dust depletion, then their \4he problem is entirely analogous 
to the one we confront in using the D abundance data, with its 
errors, to infer the {\it maximum}, gas phase D abundance.

Here, the FUSE set of Galactic ISM deuterium observations is analyzed 
assuming only that there exists a ``true", uniform, ISM deuterium 
abundance whose {\it gas phase} value may have been reduced by the 
depletion of D onto dust grains,  with the amount of depletion 
possibly varying from LOS to LOS.  The a priori unknown distribution 
of depletions is characterized in terms of a Bayesian probability 
distribution (a ``prior"), and the data themselves are used to 
determine both the true ISM abundance \ydism~= \ydmax~or, a lower 
bound to it, \ydism~$\geq$ \ydmax, as well as a measure of the amount 
of depletion, the depletion parameter $w \equiv y_{\rm D,max} - 
y_{\rm D,min}$.  No prior assumptions about which LOS may have 
been affected by dust depletion are imposed and the entire data 
set is analyzed in an unbiased, Bayesian manner.  

\subsection{The Data}
\label{subsection:data}

In our analysis we use the FUSE ISM deuterium abundance data 
for 46 LOS from \citet{linsky}, together with three, more recent 
measurements towards HD41161, HD53975 \citep{oliveira2006} and 
REJ1738+665 \citep{dupuis2009}.  Of the 49 LOS with deuterium 
abundance measurements, 21 are within the LB (see \citet{linsky} 
for a discussion of the LB); the remaining 28 are nLB LOS, 
towards stars beyond the LB.  While the star 31 Com is more 
distant than all the LB stars, the LOS to it has a very low 
\hi~column density and an average \hi~volume density ($n($\hi)$ 
\equiv $N(\hi)/$d$) much smaller than for all LB LOS.  According 
to \citet{piskunov} and \citet{dring}, the absorption feature toward 
31 Com is at a velocity which is inconsistent with the LB and 
this LOS likely lies within the hot, ionized, Local Bubble 
(which may account for the ``high" D and Fe abundances) in the 
direction of the north galactic pole.  For a contrary point of 
view, see \cite{redfield08}.  In contrast to the assignment in 
Table 2 of \citet{linsky}, we include 31 Com along with the nLB 
LOS (see Figure \ref{fig:lgyvslgh}).  However, this assignment 
has negligible impact on our quantitative results.  \citet{linsky} 
also list D/H ratios for the nLB LOS {\it corrected} by them 
for assumed {\it average} foreground LB \di~and \hi~absorption.  
If velocity information were available it could be used to 
separate foreground absorption in the LB from that due to gas 
lying beyond the LB.  In the absence of such data, \citet{linsky} 
{\it assume} that the LB extends to N(\hi)$_{\rm LB}$ $= 10^{19.2}$ 
and multiply N(\hi)$_{\rm LB}$ by an adopted average LB D/H ratio 
to find N(\di)$_{\rm LB}$.  These two average column densities 
are subtracted from the \di~and \hi~column densities observed 
for the individual nLB LOS and the ratio of the corrected 
column densities is used to find corrected nLB D/H ratios.  
\citet{linsky} base their assumed LB \hi~column density on the 
Na\thinspace{$\scriptstyle{\rm I}$} observations of \cite{lallement03}, 
who find tentative evidence for a LB "wall" of cold dense 
gas with N(\hi) $\sim 10^{19.5}$.  This procedure can bias 
the ``corrected" nLB deuterium abundances.  For example, 
it enhances the D abundances along those nLB LOS where the 
observed D abundance exceeds the adopted LB D abundance 
and decreases the D abundances for those LOS where the 
observed D abundance is less than the adopted LB value.  
The magnitude of the correction increases with the value 
of the ``average" \hi~column density adopted.  In fact, the 
observed LB \hi~column and volume densities are distributed 
very inhomogeneously, varying by nearly a factor of 30 
along different LB LOS.  This results in a scatter plot 
for the column densities as a function of distance to 
the background star.  Even for the subset of the most 
distant LB stars with $d \sim 70 - 80$~pc, N(\hi) varies by 
a factor of $\sim 16$, $10^{17.9} \la$~N(\hi) $\la 10^{19.1}$.  
The structure of the LB is very complex \citep{lallement03}
and the location of the LB "wall" varies from 65-150 pc.  
Thus, the value for the LB \hi~column density adopted by 
\citet{linsky} may be an {\it overestimate}, leading to an 
{\it overestimate} for their inferred value of \ydmax, since 
a smaller choice for the average foreground LB \hi~column 
density will result in a smaller correction to N(\di) and 
a lower estimate of \ydmax.  For these reasons, in our 
analysis, we use the uncorrected \hi~and \di~column 
densities for all nLB LOS.

Following \citet{linsky}, we adopt as a working hypothesis that 
the large scatter in the observed, gas phase, ISM deuterium 
abundances is a reflection of the preferential depletion of 
deuterium (relative to hydrogen) onto dust grains.  Therefore, 
it is assumed that in the ISM deuterium has a total LOS column 
density N(D)$_{\rm ISM}$ = N(D)$_{\rm gas}$ + N(D)$_{\rm dust}$, 
where the sum includes the observed, gas phase component, 
N(D)$_{\rm gas}$ = N(\di)$_{\rm OBS}$, and an unobserved, 
dust-depleted component, N(D)$_{\rm dust}$.  Along any LOS, 
N(D)$_{\rm ISM} \geq$ N(D)$_{\rm gas}$.  The data suggest, 
and we assume, that hydrogen is negligibly depleted onto 
dust and that the fraction of H tied up in molecules in 
the diffuse ISM probed by FUSE is usually negligible (when 
H$_{2}$ is observed, its column density is included in the 
budget for gas phase hydrogen), so that N(H)$_{\rm ISM} = 
$~N(H)$_{\rm gas}$.  

On the assumption of a uniform D abundance (gas plus dust) in 
the local ISM, the dispersion among the observed gas phase D 
abundances, $y_{{\rm D},i}$, reflects the observational errors 
in N(\di) and N(\hi), along with any spatial variations in D 
depletion.  Since deuterium may be depleted along ALL local 
ISM LOS, $y_{{\rm D},i} \leq y_{\rm D,max}$, where $y_{\rm 
D,max}$ is the maximum gas phase (undepleted) deuterium 
abundance, consistent with the observational errors, and 
\ydism~$\geq$~\ydmax.  Within the local ISM there will be a 
maximum depletion so that $y_{{\rm D},i} \geq y_{\rm D,min}$.  
Along the $i$th LOS, the observed, gas phase D abundance, 
$y_{{\rm D},i}$ ($y_{{\rm D},i} \equiv y_{{\rm D},i,{\rm 
obs}}$), may differ from the maximum undepleted abundance 
by an amount $w_{i}$ where, $w_{i} \equiv y_{\rm D,max} - 
y_{{\rm D},i}$, and $0 \leq w_{i} \leq w \equiv y_{\rm D,max} 
- y_{\rm D,min}$.  In the absence of observational errors, 
the observed D abundance along the $i$th LOS would lie 
between \ydmin~and \ydmax~and would reflect the difference 
between the maximum undepleted ISM abundance \ydmax, and 
the spatially varying value of $w$, $w_{i} = y_{\rm D,max} 
- y_{{\rm D},i}$.  Observational errors complicate the task 
of using the data, \{$y_{{\rm D},i}$\}, to identify \ydmax~(and 
\ydmin~or, $w$).

Since real data do have errors, the observationally inferred D 
abundance along the $i$th LOS, $y_{{\rm D},i}$, will differ from 
the true gas phase D abundance, $y_{{\rm D},i,\rm T}$, by an 
amount $\delta_{i}$ ($y_{{\rm D},i} = y_{{\rm D},i,\rm T} + 
\delta_{i}$).  For an individual measurement the difference 
between the observed and true D (gas phase) abundances, 
$\delta_i$, is unknown.  If it is assumed that $\delta_{i}$ is a 
random variable drawn from a zero-mean Gaussian of width 
$\sigma_i$, then the probability distribution for $y_{{\rm D},i}$, 
given a true gas phase abundance $y_{{\rm D},i,\rm T}$ along 
the $i$th LOS is
\beq
P(y_{{\rm D},i}|y_{{\rm D},i,\rm T}) = \frac{1}{\sqrt{2\pi}\sigma_{i\pm}} 
e^{-\left(y_{{\rm D},i} - y_{{\rm D},i,\rm T}\right)^2/2\sigma_{i \pm}^2}
\label{eq:gauss}
\eeq
In eq.~(\ref{eq:gauss}) we allow for asymmetrical measurement 
errors so that $\sigma_{i+}$ corresponds to $y_{{\rm D},i} > 
y_{{\rm D},i,\rm T}$ and $\sigma_{i-}$ to $y_{{\rm D},i} < 
y_{{\rm D},i,\rm T}$. 
 
\subsection{Bayesian Formalism}

For a set of data with errors \{$y_{{\rm D},i},\delta_{i}$\}, 
the Bayesian approach described by \citet{hogan} enables a 
determination of two parameters, \ydmax~and $w$ (or \ydmin).  
Along each LOS in the local ISM the true abundance is 
related to \ydmax~by $y_{{\rm D},i,\rm T} = y_{\rm D,max} 
- w_i$.

The likelihood of finding particular values of \ydmax~and of $w$ 
(or \ydmin), $\mathscr{L}(y_{\rm D,max}; w)$, from a sample of 
measurements $y_{{\rm D},i}$, with errors $\delta_{i}$ is,
\beqar
\mathscr{L}(y_{\rm D,max}; w) &=& \prod_{i} P(y_{{\rm D},i}|y_{\rm D,max};w)	\\
\mathscr{L}(y_{\rm D,max}; w) &=& \prod_{i} \int dy_{{\rm D},i,\rm T}P(y_{{\rm 
D},i}|y_{{\rm D},i,\rm T})  \times P(y_{{\rm D},i,\rm T}|y_{\rm D,max}; w)
\label{eq:def}
\eeqar
where $P(y_{{\rm D},i}|y_{{\rm D},i,\rm T})$ is the probability 
distribution from eq.~(\ref{eq:gauss}), relating each observed 
abundance $y_{{\rm D},i}$ to its underlying true value $y_{{\rm 
D},i,\rm T}$.  $P(y_{{\rm D},i,\rm T}|y_{\rm D,max}; w)$ is the 
dust depletion probability distribution -- the probability of 
finding the true, dust depleted, LOS deuterium abundance $y_{{\rm 
D},i,\rm T}$, given the values of the maximum and minimum ISM 
deuterium abundances, $y_{\rm D,max}$ and \ydmin~= \ydmax~-- $w$ 
respectively.  Our probability distribution, $P(y_{{\rm D},i,\rm 
T}|y_{\rm D,max}; w)$, determines the distribution of the $w_{i}$.  
To avoid any a priori model dependence and to limit ourselves 
to the fewest assumptions, following \citet{hogan} we initially 
adopt the simplest form for this probability distribution -- 
a {\em top-hat} function:
\beq
 P(y_{{\rm D},i,\rm T}|y_{\rm D,max}; w) = \Biggl\{ 
\begin{array}{cc}
1/w ~, & \mbox{$y_{\rm D,min} \le y_{{\rm D},i,\rm T} 
\le y_{\rm D,max}$} \\ 0 ~~~~, &\mbox{otherwise} 
\end{array} 
\label{eq:top-hat}
\eeq
where \ydmin~= \ydmax~-- $w$.  The probability distribution is 
normalized to unity when integrated over $y_{{\rm D},i,\rm T}$.  
Integrating over the range of possible deuterium abundances 
for $N$ data points, we find the likelihood distribution for 
the maximum gas phase deuterium abundance in the local ISM, 
$y_{\rm D,max}$, and for $w$,
\beqar
\label{eq:like}
\mathscr{L}(y_{\rm D,max}; w) = \prod_i^N \frac{1}{2w}  \Biggl[ {\rm 
erf} \left( \frac{y_{{\rm D},i} - y_{\rm D,min}}{\sqrt{2} \sigma_{i+}} 
\right)  - {\rm erf} \left( \frac{y_{{\rm D},i} - y_{\rm D,max}}{\sqrt{2} 
\sigma_{i-}} \right) \Biggr] 
\eeqar
Equation~(\ref{eq:like}) is evaluated numerically for a range of 
$y_{\rm D,max}$ and $w$ (or \ydmin) values to find the combination 
of these two parameters which maximizes the likelihood.  In this 
way the set of deuterium observations, including the errors, is 
used to find the most likely values of  \ydmax~and the depletion 
parameter, $w$, avoiding any prior assumptions about which LOS 
may, or may not, be affected by depletion of deuterium onto dust 
or, about how the abundance of D may, or may not, be correlated 
with the abundance of iron or other metals.
 
In addition to adopting the top-hat depletion probability 
distribution, following \citet{hogan} we also explore the 
consequences of choosing two other, asymmetric probability 
distributions:
\beqar
\label{eq:positive} 
 &P(y_{{\rm D},i,\rm T}|y_{\rm D,max}; w) =   \Biggl\{ 
\begin{array}{cc}
2 (y_{{\rm D},i,\rm T}-y_{\rm D,min})/w^{2} ~, &\mbox{$y_{\rm 
D,min} \le y_{{\rm D},i,\rm T} \le y_{\rm D,max}$} \\
0 ~~~~~~~~~, &\mbox{otherwise} 
\end{array} 
\eeqar
\beqar
\label{eq:negative}
&P(y_{{\rm D},i,\rm T}|y_{\rm D,max}; w) =  \Biggl\{ 
\begin{array}{cc}
2(y_{\rm D,max}-y_{{\rm D},i,\rm T})/w^{2} ~, &\mbox{$y_{\rm 
D,min} \le y_{{\rm D},i,\rm T} \le y_{\rm D,max}$} \\
0 ~~~~~~~~~, &\mbox{otherwise} 
\end{array} 
\eeqar
where eq.~(\ref{eq:positive}) is a {\em positive bias} distribution, 
favoring smaller depletion of deuterium, while eq.~(\ref{eq:negative}) 
is a {\em negative bias} distribution, favoring larger D-depletion.  
At the Referee's suggestion we have also analyzed the effects of 
another probability distribution -- a bimodal, {\em {\rm M}-shaped 
bias} distribution defined in eq.~(\ref{eq:Mbias}), which favors 
both low and high depletion of deuterium, while moderate depletion 
is disfavored.  Finally, we also explored the complementary, 
{\em $\Lambda$-shaped} probability distribution presented in 
eq.~(\ref{eq:lambda}).  Both additional distributions are defined 
with respect to the midpoint $y_{\rm M}=(y_{\rm D,max}+y_{\rm 
D,min})/2 = y_{\rm D,max} - w/2$.
\beqar
\label{eq:Mbias} 
 &P(y_{{\rm D},i,\rm T}|y_{\rm D,max}; w) =    \Biggl\{ 
\begin{array}{cc}
\frac{2 (y_{\rm M} - y_{{\rm D},i,\rm T})}{w(y_{\rm M}+w-y_{\rm D,max})} ~, &\mbox{$y_{\rm 
D,min} \le y_{{\rm D},i,\rm T} \le y_{\rm M}$} \\
\frac{2 (y_{\rm M} - y_{{\rm D},i,\rm T})}{w(y_{\rm M}-y_{\rm D,max})} ~, &\mbox{$y_{\rm M} 
\le y_{{\rm D},i,\rm T} \le y_{\rm D,max}$} 
\end{array} 
\eeqar

\beqar
\label{eq:lambda} 
 &P(y_{{\rm D},i,\rm T}|y_{\rm D,max}; w) = \Biggl\{ 
\begin{array}{cc}
\frac{2 (y_{{\rm D},i,\rm T} - y_{\rm D,max} + w)}{w(y_{\rm M}+w-y_{\rm D,max})} ~, &\mbox{$y_{\rm 
D,min} \le y_{{\rm D},i,\rm T} \le y_{\rm M}$} \\
\frac{2 (y_{{\rm D},i,\rm T} - y_{\rm D,max})}{w(y_{\rm M}-y_{\rm D,max})} ~, &\mbox{$y_{\rm M} 
\le y_{{\rm D},i,\rm T} \le y_{\rm D,max}$} 
\end{array} 
\eeqar
All probability distributions are normalized to unity when 
integrated over $y_{{\rm D},i,\rm T}$. 

Our Bayesian analysis determines two parameters, the best fit 
values of the maximum and minimum deuterium abundances, \ydmax~and 
\ydmin~= \ydmax~-- $w$, compatible with the errors in the data and 
the assumption that any spread in the observed abundances, above 
and beyond that expected from the errors, is real and reflects the 
variable ISM depletion of deuterium onto dust.  \citet{hogan} were 
mainly interested in the {\it minimum} \4he abundance, Y$_{min}$, 
and we are primarily interested in the {\it maximum} ISM D abundance, 
\ydmax.  The likelihood in the \{\ydmax,$w$\} plane is maximized to 
find the best estimate of the maximum undepleted D abundance, 
providing a lower limit to the ISM D abundance (\ydism~$\geq$ \ydmax).
We note that in the absence of observational errors, the depletion 
parameter, $w$, is restricted to $0 \leq w \leq y_{\rm D,max}$, 
since \ydmax~$\geq$ \ydmin.  In Figures 2 -- 4, $w$ starts at zero 
and the dashed lines show the boundary, $w = y_{\rm D,max}$.

\subsection{Results}

\begin{figure}[t]
\epsscale{0.4}
\plotone{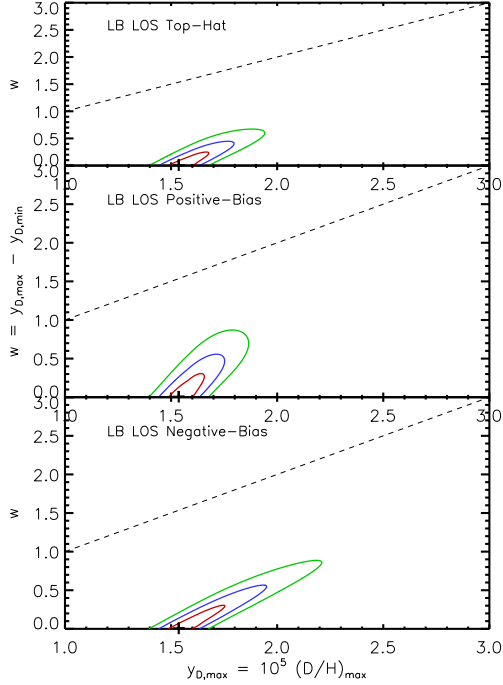}
\caption{Likelihood contours (68\%, 95\%, 99\%) in the $w$ -- 
\ydmax~plane for the 21 LB LOS for the top-hat (top panel), 
the positive-bias (middle panel), and the negative-bias 
probability distribution (bottom panel), where $w \equiv 
y_{\rm D,max} - y_{\rm D,min}$ is the depletion parameter.  
The best fit values -- for all three probability distributions 
-- are at \ydmax~= 1.5 and $w = 0$ (\ydmin~= \ydmax), rounded 
off to two significant figures.  The dashed line, $w=y_{\rm 
D,max}$, represents the bound above which the results, in 
the absence of errors, are unphysical since $y_{\rm D,min} 
= y_{\rm D,max} - w$ becomes negative.
}
\label{fig:lballpriors}
\end{figure}

As may be seen from Figure~\ref{fig:lgyvslgh}, the distributions of 
the observed D abundances for the LB and nLB LOS are very different.  
The LB \yd~values show little gas phase variation from LOS to LOS, 
in contrast to the nLB D/H ratios, whose variation accounts for 
most of the factor $\sim 4$ spread in the observed ISM D abundances.  
Indeed, all 21 LB D abundances are consistent with no variation 
(within the observational errors) around a weighted mean of 
log(\ydlb)~= 0.19, corresponding to \ydlb~= 1.5.  Our Bayesian 
analysis for all five probability distributions is in agreement 
with this and yields $w \approx 0$ and \ydmax ~$\approx$ \ydmin~$\approx 
1.5$.  Our results for three distributions (top-hat, positive-bias, 
negative-bias), shown in Figure~\ref{fig:lballpriors}, illustrate 
this result. A {\it lower} bound to the deuterium astration factor 
may be inferred from the {\it upper} bound to the LB deuterium 
abundances, \ydmax~= \ydlb, and the estimate of \ydp~from \citet{pettini}.  
For central values of log(\ydlb)~$ \geq 0.19$ and log(\ydp)~$ = 0.45$ 
\citep{pettini}, log$(f_{\rm D,LB}) \leq 0.26$, corresponding to 
$f_{\rm D,LB} \leq 1.8$~\footnote{Significant figures and round off: 
The \hi~and \di~column densities listed in Table 2 of \citet{linsky} 
are typically given to two significant figures (the integers in 
front of the decimal place simply count the powers of ten).  These 
column densities are used to find log(\yd) = 5 + log~N(\di) -- 
log~N(\hi), where the errors in the logs of the column densities 
are added in quadrature.  As a result, the values of log(\yd) 
(and their errors) are only known to {\bf two} significant figures 
and, so too are the values of \yd~(and their errors) derived from 
them.  However, in Table 3 of \citet{linsky} the values of \yd~are 
given to {\bf three} (or more) significant figures.  For our 
Bayesian analysis we use the data in Table 3 of \citet{linsky}, 
but the results presented here are generally rounded to two 
significant figures.  As an example, if we had first rounded 
the individual values of \ydlb~and \ydp~to two significant 
figures, we would have found \ydlb~= 1.5 and \ydp~= 2.8, leading 
to $f_{\rm D} \leq$~\ydp/\ydlb~= 1.9, in contrast to the slightly 
different value quoted here, which follows from log($f_{\rm D}) 
\leq$~log(\ydp) -- log(\ydlb) = 0.26 and $f_{\rm D} = 10^{0.26} 
= 1.8$.}, consistent with the successful Galactic chemical 
evolution models identified in SRT.  This value is also 
consistent with a wide range of chemical evolution models 
discussed in \citet{pf08} with both low and high infall rates 
of nearly pristine gas, as well as with a variety of initial 
mass functions.

\begin{figure}[t]
\epsscale{0.4}
\plotone{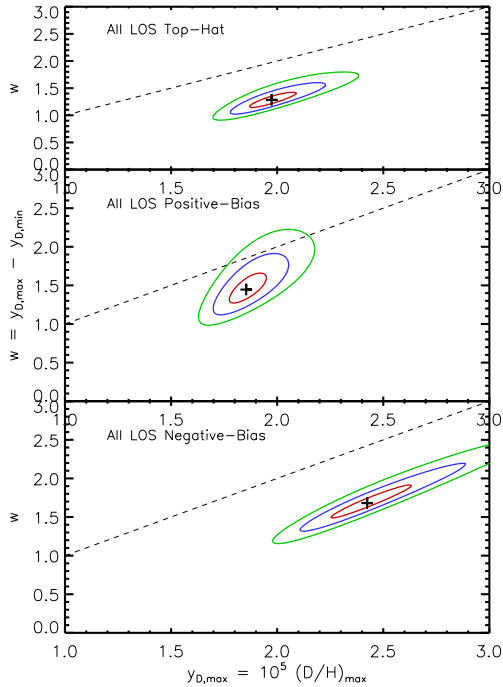}
\caption{Likelihood contours (68\%, 95\%, 99\%) in the $w$ -- 
\ydmax~plane for all 49 LOS for the top-hat (top panel), 
the positive-bias (middle panel), and the negative-bias 
probability distribution (bottom panel).  The dashed line 
is $w=y_{\rm D,max}$; see Figure~\ref{fig:lballpriors}.
}
\label{fig:allallpriors}
\end{figure}

\begin{figure}[t]
\epsscale{0.4}
\plotone{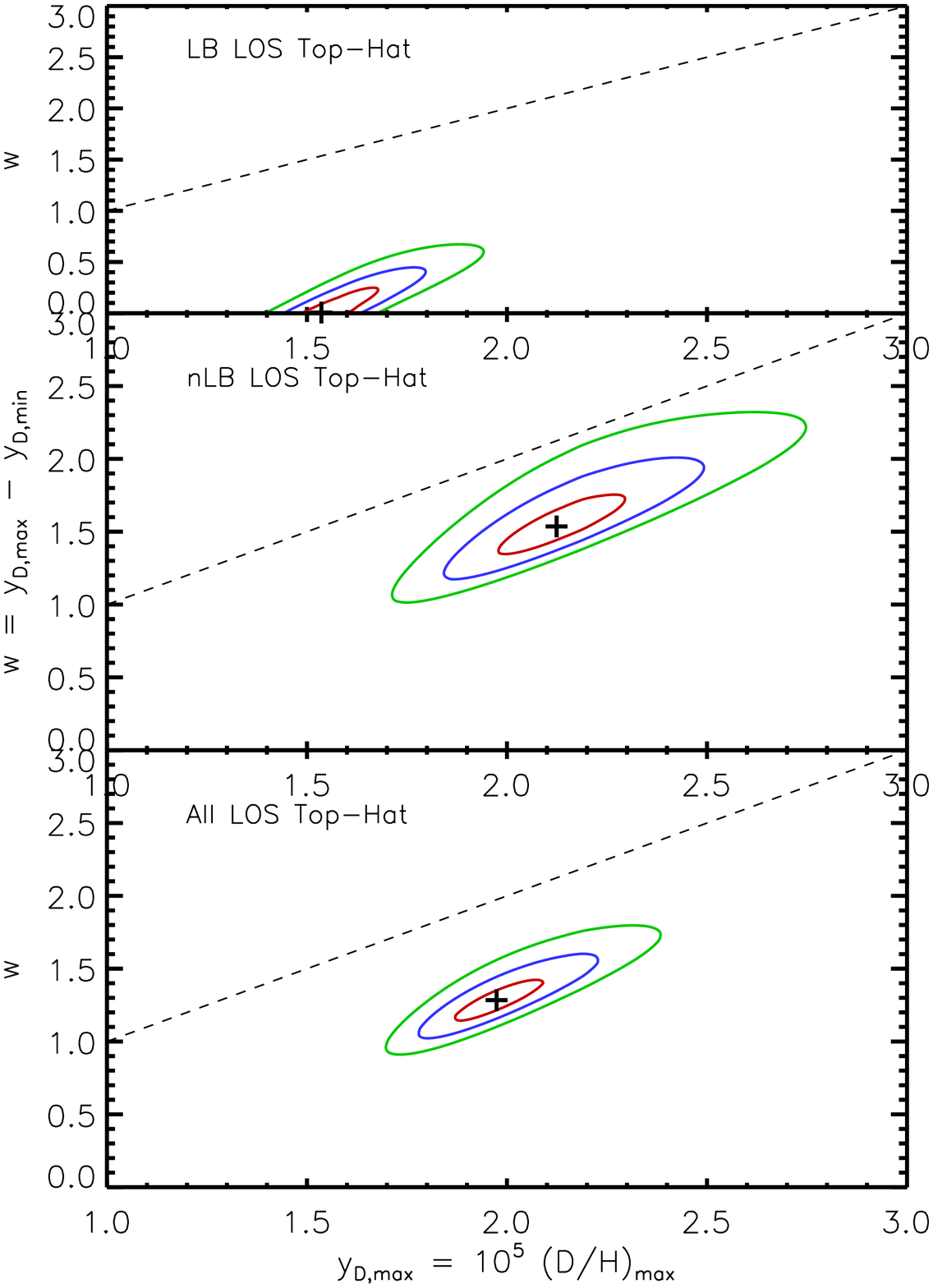}
\caption{Likelihood contours (68\%, 95\%, 99\%) in the $w$ -- 
\ydmax~plane for the 21 LB LOS (top panel), the 28 nLB LOS 
(middle panel), and all 49 LOS (bottom panel), using the 
top-hat probability distribution.  The dashed line is 
$w=y_{\rm D,max}$ (see Figure~\ref{fig:lballpriors}).
}
\label{fig:lbnlball}
\end{figure}

\begin{figure}[t]
\epsscale{0.4}
\plotone{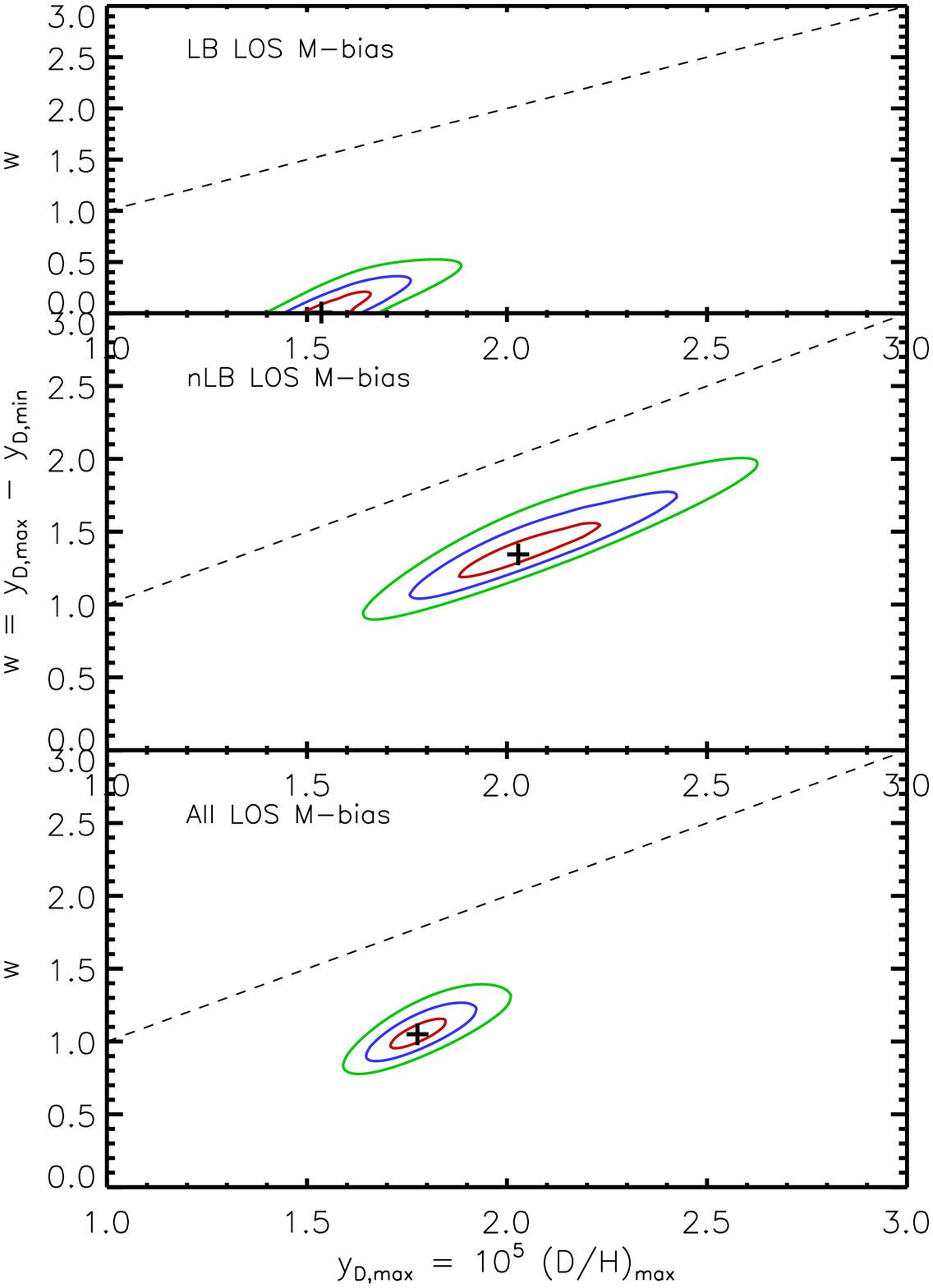}
\caption{Likelihood contours (68\%, 95\%, 99\%) in the $w$ -- 
\ydmax~plane for the 21 LB LOS (top panel), the 28 nLB LOS 
(middle panel), and all 49 LOS (bottom panel), using the 
M-shaped probability distribution.  The dashed line is 
$w=y_{\rm D,max}$ (see Figure 2).
}
\label{fig:lbnlballM}
\end{figure}

In contrast to the LB, the nLB LOS do show large variations among 
the observed gas phase D abundances (see Figure \ref{fig:lgyvslgh}).  
It is thus expected that the Bayesian analysis of this data subset 
will find evidence for $w>0$ at a statistically significant level. 
This indeed is seen in Figures~\ref{fig:lbnlball} (top-hat) and 
\ref{fig:lbnlballM} (M-shaped bias) where our results  are shown 
separately for the data of the LB (top panel), the nLB (middle 
panel) LOS, along with those for all 49 LOS (bottom panel).  For 
the 28 nLB LOS the best fit values, from a top-hat distribution, 
are \ydmax~= 2.1 and $w = 1.5$.  The top-hat Bayesian analysis for 
the complete FUSE data set (Figure~\ref{fig:lbnlball}, bottom panel) 
finds maximum likelihood values of \ydmax~= 2.0 and $w = 1.3$, 
corresponding to \ydmin~= 0.7.  The non-zero depletion parameter 
found for all 49 LOS is driven by the large variations in \yd~for 
the nLB LOS.  This value of \ydmax~corresponds to $f_{\rm D} \leq 
1.4$, which, within the errors in the measurements of \ydp~and \ydmax, 
is marginally consistent with the fiducial chemical evolution 
model discussed in SRT and is consistent with a range of models 
discussed in \citet{pf08}, where even larger infall rates are 
required for some initial mass functions.  On the other hand, 
when the complete FUSE data set is analyzed with the M-shaped 
prior suggested by the referee (Figure~\ref{fig:lbnlballM}, 
bottom panel), the maximum likelihood is obtained for \ydmax~= 
1.8 and $w = 1.0$, corresponding to \ydmin~= 0.8 and $f_{\rm D} 
\leq 1.6$.  This result is consistent with the fiducial model 
adopted in SRT, but not with some of the other Galactic chemical 
evolution models discussed by them~\citep{srt}.  Similar to the 
nLB case, this result is also consistent with some of the models 
explored in \citet{pf08}, where the observations allow for both 
high and low infall rates depending on the choice of the initial 
mass function.  However, the more recent initial mass functions 
are only consistent with larger infall rates.

As anticipated from Figure~\ref{fig:lgyvslgh}, the bottom panels 
of Figures \ref{fig:lbnlball} and \ref{fig:lbnlballM} confirm that 
the variation among the gas phase D abundances for all 49 FUSE 
LOS is too large to be accounted for by the observational errors 
($w \neq 0$ at much greater than 99\% confidence), for all adopted 
probability distributions.  Indeed, \eg for the top-hat probability 
distribution, the D abundances for all 49 LOS span a range of 
nearly a factor of three, from \ydmax~= 2.0, down to $y_{\rm D,min} 
\equiv y_{\rm D,max} - w = 0.7$.  

The effects of the choices of the priors on the likelihood 
distributions for \ydmax~and $w$ are shown for the complete 
FUSE data set on Figure \ref{fig:allallpriors} and on the 
bottom panel of Figure~\ref{fig:lbnlballM}.  While the results 
for the top-hat and positive bias distributions are quite 
similar, the negative-bias distribution is noticeably 
different.  The best fit parameters \ydmax, $w$ and resulting 
astration factor $f_{\rm D}$ are summarized in Table 1.  
Also presented in the table is, $\Delta \ln{\mathscr{L}_{\rm 
max}}$, the difference between the logarithms of the largest 
maximum likelihood and a maximum likelihood of a given 
probability distribution.  While the M-shaped probability 
distribution has the largest maximum likelihood value, 
so that $\Delta \ln{\mathscr{L}_{\rm max}} \equiv 
\ln{\mathscr{L}_{\rm max,M}} - \ln{\mathscr{L}_{\rm max,bias}}$, 
this distribution is closely followed by the top-hat and 
positive-bias priors, suggesting that $1.8 \la y_{\rm D,max} 
\la 2.0$.  The negative-bias prior, favoring large depletion, 
yields the poorest fit to the data.  As may be seen from 
Figure \ref{fig:lgyvslgh}, the large errors in the data mask 
which of our priors might provide the best fit.  Because 
the top-hat distribution requires the simplest assumption 
(no preference), and since its maximum likelihood is similar 
to that of the M-prior, we to adopt \ydmax~$= 2.0 \pm 0.1$ as 
our best estimate for the maximum of the gas phase, ISM D 
abundances.

\begin{table}
 \centering
  \caption{Results for the different shapes of the adopted priors.}
  \begin{tabular}{ccccc}
  \hline
   Bias Shape & \ydmax  & $w$ & $f_{\rm D}$   & $\Delta \ln{\mathscr{L}_{\rm max}}$  \\
 \hline
 Top-hat  & $2.0 \pm 0.1$ & $1.3 \pm 0.2$       & $1.4 \pm 0.1$  & 0.5   \\
 Positive & $1.9 \pm 0.1$ & $1.4 \pm 0.2$       & $1.5 \pm 0.1$  & 0.9   \\
 Negative & $2.4 \pm 0.2$ & $1.7^{+0.3}_{-0.2}$ & $1.2 \pm 0.1$  & 2.7   \\
 M-Shaped & $1.8 \pm 0.1$ & $1.0 \pm 0.1$       & $1.6 \pm 0.1$  & 0.0   \\
 $\Lambda$-Shaped  & $2.3^{+0.2}_{-0.1}$ & $1.8^{+0.3}_{-0.2}$ & $1.2 \pm 0.1$  & 1.4 \\
\hline
\end{tabular}
\end{table}

\section{Summary and Conclusions}

In the decades prior to the current era of precision cosmology the 
primordial abundance of deuterium provided the only quantitative 
cosmological baryometer~\citep{boes}.  Although, at present, the 
deuterium abundance is only measured along seven, high-redshift, 
low-metallicity LOS to background quasars~\citep{omeara,pettini}, 
the inferred primordial D abundance, \ydp~= $2.8  \pm 0.2$, is 
in excellent agreement with the non-BBN inferred baryon density 
parameter~\citep{steigman07}.  In the post-BBN Universe deuterium 
is destroyed as gas is cycled through stars, so that comparing 
the abundance of deuterium in the ISM of the Galaxy with the 
primordial D abundance provides an estimate of the virgin fraction 
of the ISM (\ie the amount of gas presently in the ISM which has 
never been cycled through stars), constraining models of Galactic 
chemical evolution~\citep{st92,st95,srt,pf08}.  According to 
conventional wisdom the deuterium-free, metal-enhanced products 
of stellar nucleosynthesis should be well-mixed in the local ISM.  
In contrast, the FUSE data on the abundances of deuterium and 
several metals (\eg iron, oxygen, etc.) along LOS within $\sim 1 - 2$ 
kpc of the Sun reveal a much different picture.  The FUSE~\citep{linsky} 
and earlier observations \citep{jenkins,sonneborn,hebrard,hoopes,prochaska} 
reveal unexpectedly large gas phase variations in \yd~(and in the 
abundances of iron, oxygen, etc.) within the local ISM, as shown 
for FUSE deuterium data in Figure~\ref{fig:lgyvslgh}.  It has been 
proposed that the large variations observed in the local ISM D 
abundances can be accounted for by preferential depletion of deuterium 
(relative to hydrogen) onto dust \citep{jura,draine04,draine06}, 
although incompletely mixed infall of relatively unprocessed, 
deuterium-enhanced, metal-free material may have contributed to 
some of the observed variations~\citep{st92,st95,srt}.  The large 
variations among the ISM D abundances, along with observational 
errors and the possible contributions from dust depletion and 
infall, complicate using the D observations to provide a robust 
estimate of the ISM D abundance which, in combination with the 
primordial D value, can lead to a constraint on the deuterium 
astration factor, $f_{\rm D}$.  The key question is, given the 
data (with its errors), how to find the best estimate of the 
``true", undepleted, ISM D abundance?

Here, to address this question, the limits to the true, undepleted, 
ISM D abundance were investigated employing a model-independent 
Bayesian statistical analysis similar to that used by \citet{hogan} 
to infer the primordial helium abundance from a set of helium 
abundance observations.   It was assumed, along with \citet{linsky}, 
that the spread in the observed D abundances is the result of 
incompletely homogenized D depletion onto dust in the local ISM.  
In our analysis this is modeled by five different probability 
distributions (priors) for the \yd~values.  The \yd~(actually, 
log~\yd)~values shown in Figure \ref{fig:lgyvslgh} suggest that, 
given the relatively large errors, a uniform (top-hat) distribution, 
favoring neither low-D nor high-D may be a good approximation to 
the data.  To explore the sensitivity of our result to the choice 
of the prior, we first considered two asymmetric distributions -- 
a positive-bias  prior favoring low depletion, and a negative-bias 
prior favoring large depletion, as well as two other priors -- an 
M-shaped distribution favoring both low and high depletion, and a 
complementary, $\Lambda$-shaped distribution.

Using the FUSE deuterium observations along all 49 LOS 
\citep{linsky,oliveira2006,dupuis2009}, we found the likelihoods in 
the \{\ydmax,$w$\} plane for the five choices of the Bayesian priors 
(see Figures~\ref{fig:allallpriors} and \ref{fig:lbnlballM}).  For 
all priors, the Bayesian analysis of the full data set requires 
significant depletion (\eg $w \neq 0$ at greater than 99.9\% 
confidence).  Comparing the maximum likelihood values for the 
five different distributions, we find that the bimodal, M-shaped 
distribution provides the best fit to the observed data (see 
Table 1).  However, it is important to notice that the shapes 
of the priors require an additional assumption in our analysis, 
so that the M-shaped distribution is the most model-dependent.
\footnote{The M-shaped prior favors both low and high levels of 
D depletion while strongly disfavoring intermediate depletion, 
suggesting that two competing processes may be at work: depletion 
onto dust and evaporation from dust, perhaps due to exposure to 
shocks.  To fit the M-prior scenario, both processes would have 
to be efficient and rapid to account for the deficit of intermediate 
D abundances.  The distribution of the presently available data 
(with its errors) is inconclusive and does not strongly favor 
any of the adopted prior distributions.  When more data become 
available, the Bayesian approach presented here may be used to 
learn more about the mechanism of deuterium depletion onto dust.}
  
In contrast, the top-hat prior is the least model-dependent, 
favoring all levels of depletion equally.  Given our ignorance 
of the detailed depletion mechanisms responsible for the 
observed scatter in the gas phase ISM deuterium abundances, 
we prefer to adopt for our estimate of the undepleted, ISM 
deuterium abundance, the result of the simplest, top-hat prior, 
whose maximum likelihood is similar to that of the best-fitting 
M-distribution,
\beq
y_{\rm D,ISM} \geq y_{\rm D,max} = 2.0 \pm 0.1 = 2.0(1 \pm 0.05)
\eeq  
This value is our best estimate of the true ISM D abundance 
based on the available deuterium observations in the local ISM 
and is independent of any model-dependent assumptions about 
galactic chemical evolution.  Combining our result with \ydp~= 
$2.8 \pm 0.2 = 2.8(1 \pm 0.07)$ \citep{pettini} (which, recall, 
provides a {\it lower} bound to the primordial abundance), 
yields a limit to the deuterium astration factor $f_{\rm D} 
\leq 1.4 \pm 0.1$ (for the M-prior, $f_{\rm D} \leq 1.6 \pm 
0.1$), consistent with most, but not all, Galactic chemical 
evolution models \citep{srt,pf08,romano09}.  If, on the other 
hand we compared this \ydism~value to the BBN + WMAP inferred 
primordial D abundance, \eg \ydp~= $2.5 \pm 0.1$ \citep{steigman10}, 
\ydp~= $2.5 \pm 0.2$ \citep{cyburt08} or, the prediction inferred 
when including the WMAP-determined effective number of neutrino 
species \citep{komatsu10}, \ydp~= $3.0 \pm 0.4$, the resulting 
deuterium astration factor would be somewhat lower in first 
two cases, $f_{\rm D} \approx 1.3 \pm 0.1$, which is marginally 
problematic for some GCE models.  In contrast, for the 
\citet{komatsu10} value of \ydp, $f_{\rm D} \leq 1.5 \pm 0.2$, 
which is entirely consistent with GCE models.

As seen in Figures~\ref{fig:lgyvslgh} -- \ref{fig:lbnlball}, 
for the LB there is little scatter among the gas-phase D 
abundances.  The small scatter is entirely consistent with 
the observational errors ($w = 0$) and all LB D abundances 
are consistent, within the errors, with \ydlb~= $1.5(1 \pm 
0.03)$.  This suggests that for the Local Bubble, \ydism~$\geq$ 
\ydlb~and, $f_{\rm D,LB} \leq 1.8 \pm 0.1$, consistent with 
all the successful chemical evolution models identified in 
SRT.  However, while the uniform LB D abundance suggests 
that D may be {\it undepleted} in the LB, for all LOS, 
\ydmax~$\approx 1.3$ \ydlb, suggesting either that D is {\it 
depleted uniformly} in the LB or, that outside of the LB 
the gas phase deuterium abundance may have been {\it enhanced} 
along some LOS by the addition of nearly primordial gas 
which has recently fallen into the disk of the Galaxy in 
the form of cloudlets which take some time to mix with 
the pre-existing gas in the ISM.  Does \ydlb~= 1.5 or 
\ydmax~= 2.0 provide the best estimate of the lower bound to 
the ISM D abundance?  If deuterium is depleted onto dust, 
why is there not a strong correlation between deuterium 
abundance and iron depletion\footnote{As pointed out by the 
Referee, shock strength may play a role in accounting for 
the scatter observed in the correlation between the gas 
phase D and Fe abundances.  If deuterium is loosely bound 
to the grain mantle while iron is locked into the core of 
the dust grain, deuterium would be more easily returned 
to the gas than iron when grains are exposed to shocks 
of modest strength, while iron might be removed from 
dust grains only by stronger shocks. The scatter in the 
correlation between the gas phase D and Fe abundances 
may be an indicator of shock strength.}  \citep{linsky} and, 
which refractory element is then best to use as proxy for 
determining deuterium depletion onto dust?  These questions 
cannot be answered by the analysis presented here.  In a 
companion paper \citep{sp}, abundances of refractory elements 
are used in concert with the deuterium abundances in an 
attempt to resolve this question.

\subsection*{Acknowledgments}
We are grateful to the Referee (J. L. Linsky) for a constructive 
and valuable report which has helped us improve on our original 
manuscript.  GS acknowledges valuable discussions with D. Romano 
and M. Tosi and we thank M. Tosi, C. Hogan, and V. Pavlidou for 
helpful remarks on an earlier version of this manuscript.  The 
work of TP is supported by the Provincial Secretariat for Science 
and Technological Development, and by the Ministry of Science of 
the Republic of Serbia under project number 141002B.  The research 
of GS is supported at The Ohio State University by a grant from 
the US Department of Energy.  Some of the work reported here was 
carried out when GS was a Humboldt Awardee at the Max Planck 
Institute for Physics and the Ludwig Maximillians University 
in Munich.  GS is grateful to the AvH for its support and to 
the MPI and the LMU for hospitality. 

{}

\label{lastpage}

\end{document}